\documentclass[aps,preprint]{revtex4}%
\usepackage{amsfonts}
\usepackage{amsmath}
\usepackage{amssymb}
\usepackage{graphicx}%
\providecommand{\U}[1]{\protect\rule{.1in}{.1in}}

\begin{document}
\preprint{ }
\title{Quantization of Li\'{e}nard's nonlinear harmonic oscillator and its solutions
in the framework of supersymmetric quantum mechanics}
\author{A.\ Abdellaoui$^{1,2}$}
\author{F.\ Benamira,$^{1}$}
\affiliation{$1.$ Laboratoire de Physique Th\'{e}orique, D\'{e}partement de Physique,
Facult\'{e} des Sciences Exactes, Universit\'{e} Fr\`{e}res Mentouri
Constantine 1, Route d'Ain-El-Bey, Constantine (25000), Algeria}
\affiliation{$2.$ D\'{e}partement de Physique, Facult\'{e} des Sciences Exactes,
Universit\'{e} Larbi Ben M'hidi, Oum el Bouaghi (04200), Algeria }
\begin{abstract}
Li\'{e}nard-type nonlinear one-dimensional oscillator is quantized using van
Roos symmetric ordering recipe for the kinetic-like part of the new derived
Hamiltonian. The corresponding Schr\"{o}dinger equation is exactly solved in
momuntum space via the approach of supersymmetric quantum mechanics (SUSYQM).
The bound-states energy spectra and corresponding wave functions are given
explicitly in terms of the ambiguity parameters. The limiting case of no
deformation agrees exactly with the eigenenergies and eigenfunctions of the
ordinary quantum harmonic oscillator.

\end{abstract}
\volumeyear{year}
\volumenumber{number}
\issuenumber{number}
\eid{identifier}
\date{02.07.2018}
\received[Received text]{date}

\revised[Revised text]{date}

\accepted[Accepted text]{date}

\published[Published text]{date}

\startpage{1}
\endpage{14}
\maketitle

\section{Introduction}

The study of quantum mechanical systems with a position-dependent effective
mass (PDEM) \cite{pdm} has witnessed a certain degree of importance due to
their importance in describing the physics of many microstructures and
mesoscopic structures of current interest \cite{XX}. However, a new
formulation in momentum-space to common problems starts to gain interest in
quantum mechanics \cite{momentum,chitika,lit3}.

Recently an attempt to quantify the nonlinear Li\'{e}nard-type one dimensional
differential equation%
\begin{equation}
\overset{\cdot\cdot}{x}+kx\overset{\cdot}{x}+\frac{k^{2}}{9}x^{3}+\omega
^{2}x=0;\text{\ \ \ }(\cdot\equiv\frac{d}{dt}), \label{eqn1}%
\end{equation}
defined on the real axis $(x\in%
\mathbb{R}
)$ where $k$ and $\omega$ are, \textit{a priori,} arbitrary positive
parameters, has been carried out in that space, \cite{chitika,lit3}. This
equation, which may be seen as a deformation of the linear harmonic oscillator
(\textit{lho}) equation, namely $\overset{\cdot\cdot}{x}+\omega^{2}x=0$, by
$k$-dependent terms, admits a periodic solution \cite{Lakshmanan}%

\begin{equation}
x\left(  t\right)  =\frac{A\sin\left(  \omega t+\delta\right)  }{1-\frac
{kA}{3\omega}\cos\left(  \omega t+\delta\right)  },\text{ \ }0\leq
A<\frac{3\omega}{k}, \label{eqn1-1}%
\end{equation}
that converges to the latter for $k=0$; i.e. in the absence of deformation.
For this reason, and in addition to the conditions that must satisfy any
physical solution of the Schr\"{o}dinger equation, the quantum version of
eq.(\ref{eqn1}) must coincide with the quantum harmonic oscillator in the
abesnce of deformation ($k\rightarrow0)$. A condition that is not satisfied
neither by the spectrum nor by the wave functions expressions derived in Ref.
\cite{lit3}. Indeed, though the chosen Lagrangian in Ref. \cite{lit3} leads to
eq.(\ref{eqn1}), it doesn't reduce to the \textit{lho} in the limit
$k\rightarrow0$. Consequently, the obtained results do not fulfill the
required condition in the absence of deformation.

Our goal in this work is to propose a suitable Lagrangian for eq.(\ref{eqn1})
that is not affected by this inconsistency, to deduce the corresponding
quantum Hamiltonian and to give the exact solution of the corresponding
Schr\"{o}dinger equation for the bound-states.

In section II, we use Jacobi Last Multiplier (JLM) method and its relationship
to the Lagrangian description of second order differential equations
\cite{JLM,JLM2}. Thus, a suitable Lagrangian with the corresponding classical
Hamiltonian of eq.(\ref{eqn1}) are deduced. In section III, the classical
Hamiltonian is quantized using von Roos recipe. In section IV, the
corresponding Schr\"{o}dinger equation is solved for bound-states in
momentum-space using the approach of SUSYQM after a brief review of the
latter. Then the bound-states spectrum is deduced algebrically and the
corresponding wave functions are determined. Finally, a special case
concerning the results in the limit of no deformation is discussed and the
conclusion is established.

\section{Lagrangian and Hamiltonian formulation of the Li\'{e}nard-type
equation}

The Lagrangian functions corresponding to eq.(\ref{eqn1}) may be determined by
means of the JLM method \cite{JLM,JLM2,lit3}. Indeed, one can show that for
the general second order differential equation%
\begin{equation}
\overset{\cdot\cdot}{x}+\overset{\cdot}{x}f\left(  x\right)  +g\left(
x\right)  =0, \label{eqn2-0}%
\end{equation}
particular (basic) Lagrangians are obtained as%

\begin{equation}
\widetilde{L}=\frac{1}{(2-\sigma^{-1})(1-\sigma^{-1})}\left(  \overset{\cdot
}{x}+\frac{1}{\sigma}\frac{g(x)}{f(x)}\right)  ^{2-\frac{1}{\sigma}},
\label{eqn2}%
\end{equation}
provided that $f(x)$ and $g(x)$ satisfy the condition
\begin{equation}
\frac{d}{dx}\left(  \frac{g(x)}{f(x)}\right)  =\sigma(1-\sigma
)f(x)\ \ \ \text{with}\ \ \sigma\neq0,\frac{1}{2}. \label{eqn3}%
\end{equation}

Substituting in eq.(\ref{eqn3}) $f(x)=kx$\ and $g\left(  x\right)
=\frac{k^{2}}{9}x^{3}+\omega^{2}x,$ corresponding to eq.(\ref{eqn1}), it
follows that $\sigma$ takes two values: $\sigma=\frac{1}{3},\frac{2}{3},$ such
that two basic Lagrangians are possible. Using the usual properties of
Lagrange's functions \cite{Landau}, for each value of $\sigma,$ more general
Lagrangians are given by
\begin{equation}
L=\frac{K}{(2-\sigma^{-1})(1-\sigma^{-1})}\left(  \overset{\cdot}{x}+\frac
{1}{\sigma}\frac{g(x)}{f(x)}\right)  ^{2-\frac{1}{\sigma}}+\frac{d}{dt}G(x,t),
\label{eqn2-1}%
\end{equation}
where $K$ is an arbitrary nonzero constant and $G(x,t)$ is an arbitrary
function. However, to obtain a suitable Lagrangian, which is reduced to the
\textit{lho} in the limit $k\rightarrow0$ (in the absence of the deformation),
$K$ and $G(x,t)$ have to be chosen judiciously.\ 

In this work, we will treat the case with $\sigma=\frac{2}{3}$ and set
$K=\sqrt{\frac{27\omega^{6}}{8k^{3}}}$ and $G(x,t)=$ $\frac{9\omega^{4}}%
{k^{2}}t+\frac{3\omega^{2}x}{k}$, such that our Lagrangian reads%

\begin{equation}
L\left(  x,\overset{\cdot}{x}\right)  =\frac{9\omega^{4}}{k^{2}}\left[
-\left(  1+\frac{2k}{3\omega^{2}}\overset{\cdot}{x}+\frac{k^{2}x^{2}}%
{9\omega^{2}}\right)  ^{1/2}+\frac{k}{3\omega^{2}}\overset{\cdot}{x}+1\right]
, \label{eqn4}%
\end{equation}
and is reduced in the limit $k\rightarrow0$ to the \textit{lho, }as it should
be, namely $\lim_{k\rightarrow0}L\left(  x,\overset{\cdot}{x}\right)
=L^{lho}\left(  x,\overset{\cdot}{x}\right)  =\frac{1}{2}\left(
\overset{\cdot}{x}^{2}-\omega^{2}x^{2}\right)  .$

A straightforward calculation shows that for $L\left(  x,\overset{\cdot
}{x}\right)  $ to remain a real function and that the resulting Euler-Lagrange
equation is equivalent to eq.(\ref{eqn1}), one must impose the following
constraint on the phase variables,%

\begin{equation}
1+\frac{2k\overset{\cdot}{x}}{3\omega^{2}}+\frac{k^{2}x^{2}}{9\omega^{2}}>0.
\label{eqn5}%
\end{equation}

The conjugate momentum $p$ resulting from $L\left(  x,\overset{\cdot
}{x}\right)  $ is given by
\begin{equation}
p=\frac{\partial L\left(  x,\overset{\cdot}{x}\right)  }{\partial
\overset{\cdot}{x}}=\frac{3\omega^{2}}{k}\left[  1-\left(  1+\frac{2k}%
{3\omega^{2}}\overset{\cdot}{x}+\frac{k^{2}x^{2}}{9\omega^{2}}\right)
^{-1/2}\right]  , \label{eqn6}%
\end{equation}
which, by virtue of eq.(\ref{eqn5}), must satisfy the constraint%
\begin{equation}
p\leq\frac{3\omega^{2}}{k}. \label{eqn7}%
\end{equation}

Thus, the classical Hamiltonian, $H(x,p)=p\overset{\cdot}{x}-L\left(
x,\overset{\cdot}{x}\right)  ,$ associated to eq.(\ref{eqn1}), can be written
as a function of the canonical variables $x$ and $p$ as
\begin{equation}
H(x,p)=\frac{1}{2(1-\frac{k}{3\omega^{2}}p)}\text{ }p^{2}+\frac{1}{2}\left(
1-\frac{k}{3\omega^{2}}p\right)  \omega^{2}x^{2}, \label{eqn8}%
\end{equation}
which is defined for $p\leq\frac{3\omega^{2}}{k}.$ As it should be, it is
reduced in the limit $k\rightarrow0$ to the classical Hamiltonian of the
\textit{lho, }namely $\lim_{k\rightarrow0}H(x,p)=H_{lho}(x,p)=\frac{1}%
{2}\left(  p^{2}+\omega^{2}x^{2}\right)  .$

\section{Hamiltonian quantization according to von Roos recipe}

The Hamiltonian (\ref{eqn8}) is of nonstandard-type. However, it may be
written in a suitable form similar to that of a standard position-dependent
mass Hamiltonian in the form%
\begin{equation}
H(x,p)=\frac{x^{2}}{2m(p)}+U(p), \label{eqn9}%
\end{equation}
where
\begin{equation}
m(p)=\frac{1}{\omega^{2}\left(  1-\frac{k}{3\omega^{2}}p\right)  },\text{\ }
\label{eqn10}%
\end{equation}
and%
\begin{equation}
U(p)=\frac{p^{2}}{2(1-\frac{k}{3\omega^{2}}p)}. \label{eqn11}%
\end{equation}

Hence, by interchanging the roles of the canonical variables $x$ and $p$,
$U\left(  p\right)  $ may be seen as the potential energy and $\frac{x^{2}%
}{2m\left(  p\right)  }$ as the kinetic-like part with a $p$-dependent mass
function $m(p)$. So, $H(x,p)$ looks like a Hamiltonian with position-dependent
mass, defined on the interval $p\leq\frac{3\omega^{2}}{k}$.

In order to be quantized one and obtain a Hermitian operator, we first make
use of von Roos recipe \cite{ROOS} to the kinetic-like part and write it in
the following symmetrical form
\begin{equation}
\left\{  \frac{x^{2}}{2m(p)}\right\}  =\frac{1}{4}\left[  m^{\alpha}(p)\text{
}x\text{ }m^{\beta}(p)\text{ }x\text{ }m^{\gamma}(p)+m^{\gamma}(p)\text{
}x\text{ }m^{\beta}(p)\text{ }x\text{ }m^{\alpha}(p)\right]  , \label{eqn12}%
\end{equation}
where $\alpha$, $\beta$ and $\gamma$, called ambiguity parameters, are real
and satisfy the condition $\alpha+\beta+\gamma=-1$. Hence, considering the
quantization in $p$-representation, i.e. $\left[  \widehat{x},\widehat{p}%
\right]  =i\hbar$, with $\widehat{p}\equiv p$ and $\widehat{x}=i\hbar\frac
{d}{dp}$, and after some algebraic manipulations, the quantized Hamiltonian
associated to the classical form (eq.(\ref{eqn9})) may be put in the form
\begin{equation}
H=-\frac{\hbar^{2}}{2}\frac{d}{dp}\frac{1}{m(p)}\frac{d}{dp}+V(p),
\label{eqn13}%
\end{equation}
where $V(p),$ that we call the effective potential, expresses in terms of two
free ambiguity parameters ($\alpha$ and $\gamma$)$\ $and also of some mass
terms in the form
\begin{equation}
V(p)=U(p)+\frac{\hbar^{2}}{2}\left[  \alpha\gamma\frac{(m^{\prime}(p))^{2}%
}{m^{3}(p)}+(\alpha+\gamma)\left(  \frac{(m^{\prime}(p))^{2}}{m^{3}(p)}%
-\frac{m^{\prime\prime}(p)}{2m^{2}(p)}\right)  \right]  , \label{eqn14}%
\end{equation}
with ($^{\prime}$) denotes differentiation with respect to $p$ ($^{\prime
}(=\frac{d}{dp})$). Substituting eqs.(\ref{eqn10}) and (\ref{eqn11}) into
eqs.(\ref{eqn14}) and (\ref{eqn13}), one obtains%
\begin{equation}
V(p)=\frac{1}{2\left(  1-\frac{k}{3\omega^{2}}p\right)  }\left[  p^{2}%
+\alpha\gamma\left(  \frac{\hbar k}{3\omega}\right)  ^{2}\right]  \text{ for
}p\leq\frac{3\omega^{2}}{k}, \label{eqn14-1}%
\end{equation}
and consequently,
\begin{equation}
H=-\frac{(\hbar\omega)^{2}}{2}\frac{d}{dp}\left(  1-\frac{k}{3\omega^{2}%
}p\right)  \frac{d}{dp}+\frac{1}{2\left(  1-\frac{k}{3\omega^{2}}p\right)
}\left[  p^{2}+\alpha\gamma\left(  \frac{\hbar k}{3\omega}\right)
^{2}\right]  \text{ for }p\leq\frac{3\omega^{2}}{k}. \label{eqn15}%
\end{equation}

This is our proposed quantum Hermitian Hamiltonian in momentum representation
corresponding to eq.(\ref{eqn1}).

\section{SUSYQM\ approach and exact solution of the Schr\"{o}dinger equation}

\subsection{Brief review of SUSYQM approach for PDEM Hamiltonians}

In connection to the approach of SUSYQM for constant mass supersymmetric
Hamiltonians \cite{witten1,witten2,susy}, the formalism can be extended to
PDEM Hamiltonians \cite{Bagchi-Tkat}. Considering the general Hermitian
Hamiltonian (\ref{eqn13}), one defines two associated intertwined partner
Hamiltonians, $H_{-}$ and $H_{+},$ as
\begin{equation}
H_{-}=A^{+}A, \label{eqn16}%
\end{equation}
and
\begin{equation}
H_{+}=AA^{+}, \label{eqn17}%
\end{equation}
where the adjoint operators $A$ and $A^{+}=A^{\dagger}$ are defined in terms
of the mass function $m(p)$ and the superpotential $W(p)$ as
\begin{equation}
A=\frac{\hbar}{\sqrt{2}}\frac{1}{\sqrt{m(p)}}\frac{d}{dp}+W(p),\text{
\ \ \ \ \ }A^{+}=-\frac{\hbar}{\sqrt{2}}\frac{d}{dp}\frac{1}{\sqrt{m(p)}%
}+W(p). \label{eqn18}%
\end{equation}

The partner Hamiltonians $H_{\mp}$ may be put in the following forms
\begin{equation}
H_{-}=-\frac{\hbar^{2}}{2}\frac{d}{dp}\frac{1}{m(p)}\frac{d}{dp}+V_{-}(p),
\label{eqn19}%
\end{equation}
and%
\begin{equation}
H_{+}=-\frac{\hbar^{2}}{2}\frac{d}{dp}\frac{1}{m(p)}\frac{d}{dp}+V_{+}(p),
\label{eqn20}%
\end{equation}
where the effective partner potentials $V_{\mp}(p)$ are given by
\begin{equation}
V_{-}(p)=W^{2}(p)-\frac{\hbar}{\sqrt{2}}\left(  \frac{W(p)}{m^{1/2}%
(p)}\right)  ^{\prime}, \label{eqn21}%
\end{equation}
and%
\begin{align}
V_{+}(p)  &  =W^{2}(p)+\frac{\hbar}{\sqrt{2}}\left(  \frac{W^{\prime}%
(p)}{m^{1/2}(p)}+\frac{W(p)m^{\prime}(p)}{2m^{3/2}(p)}\right)  -\frac
{\hbar^{2}}{2}\left(  \frac{3}{4}\frac{m^{\prime2}(p)}{m^{3}(p)}-\frac{1}%
{2}\frac{m^{^{\prime\prime}}(p)}{m^{2}(p)}\right)  .\nonumber\\
&  \label{eqn22}%
\end{align}

Denoting by $\varepsilon_{n}^{\mp}$ and $\psi_{n}^{\mp}\left(  p\right)  $
respectively the bound-states eigenvalues and eigenfunctions of $H_{\mp}$,
namely
\begin{equation}
H_{-}\psi_{n}^{-}\left(  p\right)  =A^{+}A\psi_{n}^{-}\left(  p\right)
=\varepsilon_{n}^{-}\psi_{n}^{-}\left(  p\right)  , \label{eqn24}%
\end{equation}
and%
\begin{equation}
H_{+}\psi_{n}^{+}\left(  p\right)  =AA^{+}\psi_{n}^{+}\left(  p\right)
=\varepsilon_{n}^{+}\psi_{n}^{+}\left(  p\right)  , \label{eqn24-1}%
\end{equation}
it follows that the spectra are semi-positive definite, i.e. $\varepsilon
_{n}^{\mp}\geq0$. Furthermore, by setting $\varepsilon_{n}^{-}=0,$ the
ground-state eigenfunction $\psi_{0}^{-}\left(  p\right)  $ is given in terms
of the superpotential $W\left(  p\right)  $ by%
\begin{equation}
\psi_{0}^{-}\left(  p\right)  =N_{0}\exp\left(  -\frac{\sqrt{2}}{\hbar}%
\int^{p}\sqrt{m\left(  p^{\prime}\right)  }W(p^{\prime})dp^{\prime}\right)  ,
\label{eqn25}%
\end{equation}
where $N_{0}$ is the normalisation constant. When the symmetry is not
spontaneously broken, i.e. if $\psi_{0}^{-}\left(  p\right)  $ is square
integrable on the domain of $p$, the eigenvalues and the corresponding
normalized eigenfunctions for all physical states of the partner Hamiltonians
are linked by%
\begin{equation}
\varepsilon_{n}^{+}=\varepsilon_{n+1}^{-}>0,\text{ for }n=0,1,2,\cdots,
\label{eqn25-1}%
\end{equation}
and%
\begin{equation}
\psi_{n}^{+}\left(  p\right)  =\frac{1}{\sqrt{\varepsilon_{n+1}^{-}}}A\psi
_{n}^{-}\left(  p\right)  ,\text{ with }n=0,1,\cdots. \label{eqn25-2}%
\end{equation}

The partner potentials $V_{\pm}(p)$ are said to be shape-invariant potentials
if they satisfy \cite{shapeinvariance}%
\begin{equation}
V_{+}(p;\left\{  \widehat{a}_{1}\right\}  )=V_{-}(p;\left\{  \widehat{a}%
_{2}\right\}  )+R\left(  \left\{  \widehat{a}_{1}\right\}  \right)  ,
\label{eqn23}%
\end{equation}
where $\left\{  \widehat{a}_{1}\right\}  $ and $\left\{  \widehat{a}%
_{2}\right\}  $ are two sets of real parameters related by a certain function
$\left(  \left\{  \widehat{a}_{2}\right\}  =f\left(  \left\{  \widehat{a}%
_{1}\right\}  \right)  \right)  $ and the remainder function $R\left(
\left\{  \widehat{a}_{1}\right\}  \right)  $ is independent of $p$.

If the requirement (\ref{eqn23}) is satisfied, the full energy spectrum
$\varepsilon_{n}^{-}$ can be deduced algebraically
\cite{shapeinvariance,witten2,susy}:%
\begin{equation}
\varepsilon_{0}^{-}=0\text{ , \ \ \ }\varepsilon_{n}^{-}=\sum_{i=1}%
^{n}R\left(  \left\{  \widehat{a}_{i}\right\}  \right)  \text{ for
}n=1,2,\cdots, \label{eqn26}%
\end{equation}
where $\left\{  \widehat{a}_{i}\right\}  =\underset{\left(  i-1\right)  \text{
times}}{\underbrace{f\circ f\circ\ldots\circ f}}\left(  \left\{
\widehat{a}_{1}\right\}  \right)  $.

In addition, the normalized excited states eigenfunctions are given by the
recurrence formula \cite{recurence}:%
\begin{equation}
\psi_{n}^{-}\left(  p;\left\{  \widehat{a}_{1}\right\}  \right)  =A^{+}\left(
p;\left\{  \widehat{a}_{1}\right\}  \right)  \psi_{n-1}^{-}\left(  p;\left\{
\widehat{a}_{2}\right\}  \right)  \text{ for }n=1,2,\ldots. \label{eqn28}%
\end{equation}

\subsection{Energy spectrum and corresponding wave functions}

Our goal is to solve the time-independent Schr\"{o}dinger equation associted
to the derived Hamiltonian (\ref{eqn15}) for bound-states, using SUSYQM
approach,
\begin{equation}
H\psi_{n}\left(  p\right)  =\varepsilon_{n}\psi_{n}\left(  p\right)  .
\label{eqn29}%
\end{equation}

Then, the challenge is to find the superpotential $W\left(  p\right)  $ so
that the Hamiltonians $H$ and $H_{-}$, given respectively by eqs.(\ref{eqn15})
and (\ref{eqn19}), are related by
\begin{equation}
H-\varepsilon_{0}=H_{-}, \label{eqn30}%
\end{equation}
where $\varepsilon_{0}$ is the ground-state energy of $H.$ Otherwise, $H$ and
$H_{-}$ share the same eigenfunctions, $\psi_{n}\left(  p\right)  =\psi
_{n}^{-}\left(  p\right)  ,$ and the energy spectra are related by
\begin{equation}
\varepsilon_{n}=\varepsilon_{0}+\varepsilon_{n}^{-}. \label{eqn31}%
\end{equation}

Inserting eqs.(\ref{eqn13}), (\ref{eqn19}), (\ref{eqn21}) into eq.(\ref{eqn30}%
), the superpotential $W\left(  p\right)  $ must satisfy the following
Riccati-like nonlinear differential equation
\begin{equation}
W^{2}\left(  p\right)  -\frac{\hbar}{\sqrt{2}}\left(  \frac{W\left(  p\right)
}{\sqrt{m\left(  p\right)  }}\right)  ^{\prime}=V\left(  p\right)
-\varepsilon_{0}, \label{eqn32}%
\end{equation}
where $V(p)$ is given by eq.(\ref{eqn14-1}). We suggest the superpotential in
the form
\begin{equation}
W\left(  p;a,b\right)  =\frac{ap+b}{\sqrt{1-\frac{k}{3\omega^{2}}p}},
\label{eqn33}%
\end{equation}
depending on two real parameters $a$ and $b,$ to be fixed in such way that
satisfy eq.(\ref{eqn32}) and so that the resulting normalized ground-state
eigenfunction (eq.(\ref{eqn25})):%
\begin{equation}
\psi_{0}\left(  p;a,b\right)  =N_{0}\left(  1-\frac{k}{3\omega^{2}}p\right)
^{\frac{3\sqrt{2}\omega}{\hbar k}\left(  b+\frac{3\omega^{2}}{k}a\right)
}\exp\left(  \frac{3\sqrt{2}\omega a}{\hbar k}p\right)  , \label{eqn34}%
\end{equation}
is square integrable in the interval $p\leq\frac{3\omega^{2}}{k}.$

To satisfy the latter requirement, $\psi_{0}\left(  p;a,b\right)  $ must
verify the boundary conditions $\psi_{0}\left(  -\infty;a,b\right)  =$
$\psi_{0}\left(  \frac{3\omega^{2}}{k};a,b\right)  =0$, which require that
$a>0$ and $b>-\frac{3\omega^{2}}{k}a$. On the other hand, substituting
eqs.(\ref{eqn33}) and (\ref{eqn14-1}) into eq.(\ref{eqn32}), it is
straightforward to deduce that the parameters $a$ and $b$ are given by
\begin{equation}
a=\frac{1}{\sqrt{2}},\text{ \ \ }b=\frac{\hbar k}{3\sqrt{2}\omega}\left(
\lambda-\mathsf{a}\right)  , \label{eqn35}%
\end{equation}
where we use the notations%
\begin{equation}
\lambda=\sqrt{\mathsf{a}^{2}+\alpha\gamma}\text{ \ \ \ \ and \ \ \ }%
\mathsf{a}=(\frac{9\omega^{3}}{\hbar k^{2}}), \label{eqn36}%
\end{equation}
with the following constraint on the ambiguity parameters%
\begin{equation}
\alpha\gamma\geq-\text{ }\mathsf{a}^{2}. \label{eqn37}%
\end{equation}

Consequently, the resulting ground-state energy, $\varepsilon_{0},$ that we
obtain from eq.(\ref{eqn32}), is given by
\begin{equation}
\varepsilon_{0}=\left(  \frac{1}{2}+\lambda-\mathsf{a}\right)  \hbar\omega.
\label{eqn38}%
\end{equation}

\subsubsection{Energy spectrum $\varepsilon_{n}$}

Inserting eqs.(\ref{eqn10}) and (\ref{eqn33}) into eqs.(\ref{eqn21}),
(\ref{eqn22}), and after some algebra, one can show that $V_{\mp}(p;a,b)$ may
be put in the compact forms
\begin{equation}
V_{-}(p;a,b)=\frac{\left(  ap+b\right)  ^{2}}{(1-\frac{k}{3\omega^{2}}%
p)}-\frac{\hbar\omega}{\sqrt{2}}a, \label{eqn39}%
\end{equation}
and%
\begin{equation}
V_{+}(p;a,b)=\frac{\left(  ap+b+\frac{\hbar k}{6\sqrt{2}\omega}\right)  ^{2}%
}{(1-\frac{k}{3\omega^{2}}p)}+\frac{\hbar\omega}{\sqrt{2}}a. \label{eqn40}%
\end{equation}
Thus, taking $a_{1}=a_{2}=a$, $b_{1}=b$ and $b_{2}=b_{1}+$ $\frac{\hbar
k}{6\sqrt{2}\omega},$ one can be easily convinced that the shape invariance
condition (\ref{eqn23}) is satisfied for the partner $V_{\mp}(p;a,b)$ and that
the remainder function depend only on the parameter $a$ as
\begin{equation}
R(a_{1},b_{1})=\sqrt{2}a\hbar\omega. \label{eqn41}%
\end{equation}
Thus, combining eqs.(\ref{eqn26}), (\ref{eqn31}), (\ref{eqn35}) and
(\ref{eqn38}) the spectrum is given by%
\begin{equation}
\varepsilon_{n}=\left(  n+\frac{1}{2}+\lambda-\mathsf{a}\right)  \hbar
\omega\text{ for }n=0,1,2,\cdots. \label{eqn42}%
\end{equation}

Note that the eigenenergies $\varepsilon_{n}$ differ from those of the quantum
harmonic oscillator only by a constant shift $\left(  \lambda-\mathsf{a}%
\right)  $\ that is depending explicitly on deformation and ambiguity
parameters. Obviously, this shift is not important since it can be absorbed in
the definition of the quantum Hamiltonian. What is important physically is
that the energy levels are equidistant with a width%
\begin{equation}
\Delta\varepsilon_{n}=\varepsilon_{n+1}-\varepsilon_{n}=\hbar\omega,
\end{equation}
just like the nondeformed linear quantum oscillator. This of course means that
neither deformation nor von Roos symmetrisation modifies the physical
character of the problem, irrespective of the choice of deformation and
ambiguity parameters.

In addition, the shift is zero in case $k\rightarrow0$ (no deformation)
whatever the choice of the ambiguity parameters $\alpha,\gamma$ and is also
zero for $\alpha\gamma=0$ whatever the choice of the deformation parameter
$k$:
\begin{equation}
\lim_{k\rightarrow0}\varepsilon_{n}=\left.  \varepsilon_{n}\right\vert
_{\alpha\gamma=0}=(n+\frac{1}{2})\hbar\omega. \label{eqn42-1}%
\end{equation}

\subsubsection{Wave functions $\psi_{n}(p)$}

While the eigenfunctions $\psi_{n}(p)$ can be deduced from the recurrence
formula eq.(\ref{eqn28}), we shall deduce them by direct calculation. By using
the variable change $y=2\mathsf{a}\left(  1-\frac{1}{\sqrt{\mathsf{a}%
\hbar\omega}}p\right)  $ where $0\leq y<\infty$, the Hamiltonian (\ref{eqn15})
may be written in a compact form as
\begin{equation}
H=-\hbar\omega\left(  y\frac{d^{2}}{dy^{2}}+\frac{d}{dy}-\frac{\lambda^{2}}%
{y}-\frac{y}{4}+\mathsf{a}\right)  , \label{eqn43}%
\end{equation}
with $\mathsf{a}$ and $\lambda$ were defined previously.\ Now, setting the
normalized wave functions in the form
\begin{equation}
\psi_{n}\left(  p\right)  =N_{n}y^{\lambda}e^{-\frac{y}{2}}\varphi_{n}\left(
y\right)  \label{eqn43-1}%
\end{equation}
where $N_{n}$ are the normalisation constants, and making use of the energy
spectrum expression (\ref{eqn42}), it follows from the Schr\"{o}dinger
equation (\ref{eqn29}), after some straightforward algebra, that the new
functions $\varphi_{n}\left(  y\right)  $ satisfies the differential equation
\begin{equation}
y\varphi_{n}^{\prime\prime}\left(  y\right)  +\left(  1+2\lambda-y\right)
\varphi_{n}^{\prime}\left(  y\right)  +n\varphi_{n}\left(  y\right)  =0,
\label{eqn44}%
\end{equation}
which is only the differential equation of the associated Laguerre
polynomials, $L_{n}^{2\lambda}\left(  y\right)  $ \cite{nikorov}.
Consequently, the corresponding normalized wave functions are given by%

\begin{equation}
\psi_{n}(p;\mathsf{a},\lambda)=N_{n}\left(  2\mathsf{a}\left(  1-\frac
{p}{\sqrt{\mathsf{a}\hbar\omega}}\right)  \text{\ }\right)  ^{\lambda}%
\exp\left(  -\mathsf{a}\left(  1-\frac{p}{\sqrt{\mathsf{a}\hbar\omega}%
}\right)  \text{ }\right)  L_{n}^{2\lambda}\left(  2\mathsf{a}\left(
1-\frac{p}{\sqrt{\mathsf{a}\hbar\omega}}\right)  \right)  , \label{eqn45}%
\end{equation}
where $N_{n}$ may be straightforwardly evaluated \cite{Gradstein}$,$%
\begin{equation}
N_{n}=\sqrt{\sqrt{\frac{\mathsf{a}}{\hbar\omega}}\frac{2n!}{\Gamma\left(
2\lambda+n+1\right)  }}. \label{eqn46}%
\end{equation}

The eigenfunctions (\ref{eqn45}) explicitly depend on the deformation and
ambiguity parameters as expected. However, they express in terms of Laguerre
polynomials, which suggests that the original problem is related to the
isotonic oscillator. However, for the latter, though the energy levels are
also equidistant, their width is twice that of the linear oscillator, namely
\cite{ZHU,dlyjr2007,bfn2016}
\[
\left.  \Delta\varepsilon_{n}\right\vert _{\text{isotonic}}=2\hbar\omega.
\]

In fact, the isotonic character of the eigenfunctions comes from the
representation used to express them. Since for $k\neq0$ the momentum $p$
varies on a half axis ($p\leq\frac{3\omega^{2}}{k})$, we can not expect the
corresponding eigenfunctions to be expressed in terms of Hermite polynomials,
which are defined on the whole axis. It turns out that in momentum
representation, the eigenfunctions are expressed in terms of Laguerre
polynomials which are well defined on a half-axis. However, as we shall see
later, in the limit $k\rightarrow0$, i.e. in the absence of deformation, the
range of variation of the momentum $p$ will extend to the entire axis and
consequently the eigenfunctions are expressed in terms of Hermite polynomials.

\paragraph{Special case}

We have shown that in the limit $k\rightarrow0,$ the eigenvalues
$\varepsilon_{n}$ coincide with those of the quantum harmonic oscillator,
eq.(\ref{eqn42-1}). Let's see what it is about the corresponding
eigenfunctions. We have $\lim_{k\rightarrow0}\lambda\approx
\mathsf{a\rightarrow\infty,}$ such that in that limit the eigenfunctions are
independent of the ambiguity parameters and read%

\begin{align}
\psi_{n}^{\text{\textit{lho}}}\left(  p\right)   &  =\lim_{\mathsf{a}%
\rightarrow\infty}\psi_{n}\left(  p;\mathsf{a},\lambda\right)  =\lim
_{\mathsf{a}\rightarrow\infty}\sqrt{2\sqrt{\frac{\mathsf{a}}{\hbar\omega}}%
}\sqrt{\frac{n!}{\Gamma\left(  2\mathsf{a}+n+1\right)  }}\left(
2\mathsf{a}\left(  1-\frac{p}{\sqrt{\mathsf{a}\hbar\omega}}\right)  \right)
^{\mathsf{a}}\times\nonumber\\
&  \left\{  \exp\left(  -\mathsf{a}\left(  1-\frac{p}{\sqrt{\mathsf{a}%
\hbar\omega}}\right)  \right)  L_{n}^{2\mathsf{a}}\left(  2\mathsf{a}%
-2\sqrt{\mathsf{a}}\left(  \frac{p}{\sqrt{\hbar\omega}}\right)  \right)
\right\}  . \label{eqn50}%
\end{align}

In order to evaluate $\lim_{\mathsf{a}\rightarrow\infty}\psi_{n}\left(
p;\mathsf{a},\lambda\right)  $, we have to proceed as follows:

(\textit{i})\textit{- }Fixing $n$ while $\mathsf{a}$ $\rightarrow\infty$,
using the Gamma function functional equation (see \cite{nikorov}), we have by
recurrence:\textbf{\ }%
\begin{equation}
\Gamma\left(  2\mathsf{a}+n+1\right)  \underset{\mathsf{a}\rightarrow
\infty}{\approx}\left(  2\mathsf{a}\right)  ^{n+1}\Gamma\left(  2\mathsf{a}%
\right)  . \label{eqn51}%
\end{equation}
In addition, $\Gamma\left(  2\mathsf{a}\right)  $ may be expressed in the
asymptotic region as \cite{nikorov}:%
\begin{equation}
\ln\Gamma\left(  2\mathsf{a}\right)  =(2\mathsf{a-}\frac{1}{2})\ln
(2\mathsf{a})-2\mathsf{a+}\frac{1}{2}\ln2\pi+O\left(  \sim(2\mathsf{a}%
)^{-(2i+1)}\right)  \text{ for }i=0,1,2,\cdots, \label{eqn52}%
\end{equation}
where $O\left(  \sim(2\mathsf{a})^{-(2i+1)}\right)  $ are correction terms
that tend to zero as $\mathsf{a}$ $\rightarrow\infty$. Hence, eq.(\ref{eqn51})
leads to
\begin{equation}
\frac{1}{\Gamma\left(  2\mathsf{a}+n+1\right)  }\underset{\mathsf{a}%
\rightarrow\infty}{=}\frac{1}{\left(  2\mathsf{a}\right)  ^{n+1}}%
(\frac{\mathsf{a}}{\pi})^{1/2}(2\mathsf{a})^{-2\mathsf{a}}\exp(2\mathsf{a}).
\label{eqn53}%
\end{equation}

(\textit{ii})\textit{-} Next, the second term in eq.(\ref{eqn50}) may be
written in a more compact form as
\begin{equation}
\left(  2\mathsf{a}\left(  1-\frac{p}{\sqrt{\mathsf{a}\hbar\omega}}\right)
\right)  ^{\mathsf{a}}\exp\left(  -\mathsf{a}\left(  1-\frac{p}{\sqrt
{\mathsf{a}\hbar\omega}}\right)  \right)  =\left(  2\mathsf{a}\right)
^{\mathsf{a}}e^{-\mathsf{a}}\left(  \frac{1-\frac{p}{\sqrt{\mathsf{a}%
\hbar\omega}}}{\exp(-\frac{p}{\sqrt{\mathsf{a}\hbar\omega}})\text{\ }}\right)
^{\mathsf{a}}. \label{eqn54}%
\end{equation}
Also, for $\mathsf{a}\rightarrow\infty$, one can write%
\begin{equation}
1-\frac{p}{\sqrt{\mathsf{a}\hbar\omega}}=\exp\left(  \ln\left(  1-\frac
{p}{\sqrt{\mathsf{a}\hbar\omega}}\right)  \right)  \underset{\mathsf{a}%
\rightarrow\infty}{\approx}\exp\left(  -\frac{p}{\sqrt{\mathsf{a}\hbar\omega}%
}-\frac{1}{2}\frac{p^{2}}{\mathsf{a}\hbar\omega}\right)  , \label{54-1}%
\end{equation}
such that
\begin{equation}
\frac{1-\frac{p}{\sqrt{\mathsf{a}\hbar\omega}}}{\exp(-\frac{p}{\sqrt
{\mathsf{a}\hbar\omega}})}\underset{\mathsf{a}\rightarrow\infty}{\approx}%
\exp\left(  -\frac{p^{2}}{2\mathsf{a}\hbar\omega}\right)  , \label{eqn55}%
\end{equation}
and consequently
\begin{equation}
\left(  2\mathsf{a}\left(  1-\frac{p}{\sqrt{\mathsf{a}\hbar\omega}}\right)
\right)  ^{\mathsf{a}}\exp\left(  -\mathsf{a}\left(  1-\frac{p}{\sqrt
{\mathsf{a}\hbar\omega}}\right)  \right)  \underset{\mathsf{a}\rightarrow
\infty}{\approx}\left(  2\mathsf{a}\right)  ^{\mathsf{a}}e^{-\mathsf{a}}%
\exp\left(  -\frac{p^{2}}{2\hbar\omega}\right)  . \label{eqn56}%
\end{equation}

(\textit{iii})\textit{- }Finally, an asymptotic calculation \cite{Gradstein}
shows that the Laguerre polynomials $L_{n}^{2\mathsf{a}}\left(  2\mathsf{a}%
-2\sqrt{\mathsf{a}}\left(  \frac{p}{\sqrt{\hbar\omega}}\right)  \right)  $ may
be reduced to the Hermite polynomials\text{ }$H_{n}\left(  \frac{p}%
{\sqrt{\hbar\omega}}\right)  $ via:%

\begin{equation}
\lim_{\mathsf{a}\rightarrow\infty}(2\sqrt{\mathsf{a}})^{-n}L_{n}^{2\mathsf{a}%
}\left(  2\mathsf{a}-2\sqrt{\mathsf{a}}\left(  \frac{p}{\sqrt{\hbar\omega}%
}\right)  \right)  =\frac{1}{2^{n}n!}H_{n}\left(  \frac{p}{\sqrt{\hbar\omega}%
}\right)  , \label{eqn57}%
\end{equation}
where now, according to eq.(\ref{eqn7}), $p\in\left]  -\infty,\infty\right[
.$

Hence, combining equations (\ref{eqn53}), (\ref{eqn56}) and (\ref{eqn57}), we
deduce that $\psi_{n}^{\text{\textit{lho}}}\left(  p\right)  $\ in
eq.(\ref{eqn50}) is given by%

\begin{equation}
\psi_{n}^{\text{\textit{lho}}}\left(  p\right)  =\frac{1}{\sqrt{2^{n}%
n!\sqrt{\pi\hbar\omega}}}\exp\left(  -\frac{p^{2}}{2\hbar\omega}\right)
H_{n}\left(  \frac{p}{\sqrt{\hbar\omega}}\right)  , \label{eqn58}%
\end{equation}
which is, obviously, the eigenfunction of the quantum harmonic oscillator
corresponding to the eigenenergy (\ref{eqn42-1}).

\section{Conclusion}

In this paper, we have investigated the bound-states solutions of the quantum
version of the Li\'{e}nard-type nonlinear oscillator. We have first used JLM
method in order to obtain a suitable Lagrangian for the problem. It turned out
then that the corresponding Hamiltonian is not of the standard type since it
requires the inversion of the roles of the canonical variables to put it in a
form typical to the position-dependent mass Hamiltonians. Thus, its
quantization required the use of von Roos recipe to write it in Hermitian
form, depending on two ambiguity parameters.

To solve the corresponding Schr\"{o}dinger equation, we used the SUSYQM
approach which allowed us to deduce the spectrum in an elegant algebraic way.
The corresponding eigenfunctions are obtained directly using a direct
resolution method based on an appropriate point transformation. It turned out
then that even if the spectrum depends explicitly on the ambiguity parameters,
the latter do not modify the physics of the problem, which remains equivalent
to the quantum harmonic oscillator.

What follows from the investigation of this problem is that from the energetic
point of view, Li\'{e}nard's nonlinear oscillator remains equivalent to the
ordinary quantum harmonic oscillator since they share relatively the same
spectrum. However, its eigenfunctions in momentum representation mimic those
of the isotonic quantum oscillator.

\end{document}